\documentclass[aps,prb,preprint,groupedaddress,floatfix]{revtex4}
\usepackage{graphicx,amssymb}
\begin{document}
\title{Evaluation of the grand-canonical partition function using Expanded Wang-Landau simulations. IV. Performance of many-body force fields and tight-binding schemes for the fluid phases of Silicon.}
\author{Caroline Desgranges and Jerome Delhommelle}
\affiliation{Department of Chemistry, 151 Cornell Street Stop 9024, University of North Dakota, Grand Forks ND 58202}

\date{\today}

\begin{abstract}
We extend Expanded Wang-Landau (EWL) simulations beyond classical systems and develop the EWL method for systems modeled with a tight-binding Hamiltonian. We then apply the method to determine the partition function and thus all thermodynamic properties, including the Gibbs free energy and entropy, of the fluid phases of Si. We compare the results from quantum many-body (QMB) tight binding models, which explicitly calculate the overlap between the atomic orbitals of neighboring atoms, to those obtained with classical many-body force fields (CMB), which allow to recover the tetrahedral organization in condensed phases of Si through e.g. a repulsive 3-body term that favors the ideal tetrahedral angle. Along the vapor-liquid coexistence, between $3000$~K and $6000$~K, the densities for the two coexisting phases are found to vary significantly (by $5$ orders of magnitude for the vapor and by up to $25$~\% for the liquid) and to provide a stringent test of the models. Transitions from vapor to liquid are predicted to occur for chemical potentials that are $10-15$~\% higher for CMB models than for QMB models, and a ranking of the force fields is provided by comparing the predictions for the vapor pressure to the experimental data. QMB models also reveal the formation of a gap in the electronic density of states of the coexisting liquid at high temperatures. Subjecting Si to a nanoscopic confinement has a dramatic effect on the phase diagram, with e.g. at $6000$~K a decrease in liquid densities by about $50$~\% for both CMB and QMB models and an increase in vapor densities between $90$~\% (CMB) and $170$\% (QMB). The results presented here provide a full picture of the impact of the strategy (CMB or QMB) chosen to model many-body effects on the thermodynamic properties of the fluid phases of Si.
\end{abstract}

\maketitle

\section{Introduction}

In recent years, the development of force fields, that are able to take into account many-body effects~\cite{tao2012accurate,distasio2014many,Nasrabad2,Moosavi,Song,Malijevsky,Nasrabad3,jager2011ab,del2013analytical,guzman2011effective,cencek2013three,Sadus,wiebke2012sensitivity,tang2012long,wiebke2014can}, has been the focus of intense research. This is especially crucial since the addition of many-body terms has been shown to improve greatly the accuracy of the predictions from molecular simulation calculations for a wide range of systems, from simple systems of rare gases and their mixtures~\cite{van1999three,jakse2002effects,Wang,Moosavi,Malijevsky2,goharshadi2006molecular,Nasrabad,Leonhard,Sadus,Anta,WangSadus,Vogt,Sadus2,jctc2015}, to molecular systems~\cite{Eckl,tkatchenko2009accurate,tkatchenko2012accurate,pai2014ab,kennedy2014communication,tkatchenko2012first,bereau2014toward,Skinner,Paesani,mas2003ab,mcdaniel2014first,schmidt2015transferable}, nanostructures~\cite{gobre2013scaling,tkatchenko2015current} and biological systems~\cite{von2010two}. These effects have also been shown to become increasingly significant in highly inhomogenous systems, such as e.g. nanoconfined systems~\cite{lee1984structure,eslami2012local,reiter2014quantum,strekalova2012hydrophobic,wu2007behavior,fullerton2005molecular,habenicht2011ab,de2012relations,kalcher2010electrolytes,li2012wetting,stroberg2012hydrodynamics}. In the case of Silicon, simulations using many-body force fields have led to a new understanding of a wide range of phenomena such as e.g. the point-defect aggregation in Silicon, the nucleation and growth of Silicon crystals as well as the formation of carbon nanotubes at SiC interfaces~\cite{choudhary2005characterizing,Honda,Makhov,Lorazo,sinno2007bottom,kapur2010detailed,desgranges2011role,kuwahara2012development,gehrmann2015reduced,ogasawara2014growth}. Different strategies have been proposed to model many-body interactions in Silicon, either relying on a purely classical approach or on a quantum approach. The idea underlying the classical many-body force fields (CMB), such as e.g. the well-known Stillinger-Weber potential~\cite{SW}, consists in using a combination of a two-body potential with an effective many-body potential (e.g. a repulsive 3-body term that favors the ideal tetrahedral angle, $\cos \theta=1/3$, between triplets of Si atoms~\cite{SW}). This allows to recover the tetrahedral arrangement of Si atoms found in the condensed phases of Si. The alternative approach, used in quantum many-body (QMB) force fields, consists in evaluating the overlap between the atomic orbitals of neighboring Si atoms, as e.g. calculated in the tight-binding Hamiltonian matrix~\cite{Chadi,Sankey,DFTB,Harrison,Cohen,Menon,Elstner,Kwon,Lenosky,cook1993comparison,ohta2008rapid,ohta2009density,berdiyorov2014stabilized,cui2014density}. In this case, the tetrahedral ordering in the condensed phases of Si directly results from the overlap between the 4 valence orbitals of Si atoms. While a comparison of the CMB and QMB approaches has been made recently on the crystalline phases of Si and on Si clusters~\cite{Ghasemi}, a full assessment of the relative performance of these two classes of force fields for the fluid phases of Si and for nanoconfined Si has yet to be carried out. In this work, in order to carry out this assessment, we extend the recently developed Expanded-Wang Landau (EWL) simulations beyond classical systems~\cite{PartI,PartII,PartIII}. Given the successes of tight-binding approaches~\cite{DFTB,ohta2008rapid,ohta2009density,berdiyorov2014stabilized,cui2014density} in computational materials science, we develop the EWL formalism to study systems modeled within tight-binding schemes. The EWL approach is an accurate and versatile scheme that allows to determine the grand-canonical partition function of systems~\cite{PartI,PartII,PartIII}. This, in turn, gives a direct access to all thermodynamic properties, including the Gibbs free energy and the entropy, from the partition functions, through the application of the formalism of statistical mechanics. Using EWL simulations, we determine the thermodynamics properties of the fluid phases of Si under a wide range of conditions, i.e. at the vapor-liquid coexistence, in compressed liquids and under a nanoscopic confinement. We consider both CMB and QMB types of models. For CMB force fields, we consider the Stillinger-Weber potential (CMB-SW)~\cite{SW} and the Tersoff potential (CMB-T)~\cite{Tersoff}. For QMB force fields, we use the Kwon model (QMB-K)~\cite{Kwon} and the Lenosky model (QMB-L)~\cite{Lenosky}. Applying the EWL approach to Si, modeled with CMB or QMB force fields, provides a full picture of the impact of the two types of strategies (CMB or QMB) on the thermodynamic properties of the fluid phases of Si in a wide range of conditions and settings. The paper is organized as follows. In the next section, we discuss how we extend the EWL approach for QMB tight-binding systems. We also detail how the EWL approach is used in conjunction with CMB force fields. Then, we present the EWL results obtained, using both classes of model, for the grand-canonical partition function of Si in the bulk and under a nanoscopic confinement. In particular, we assess the relative performance of each model and carry out a comparison of the EWL results to the experimental data. We finally draw the main conclusions from this work in the last section.

\section{Expanded Wang-Landau sampling for tight-binding models}

\subsection{Theoretical framework}

In the first papers of the series~\cite{PartI,PartII,PartIII}, we have developed the Expanded Wang-Landau approach to determine the grand-canonical partition function of single-component systems and mixtures modeled with classical force fields. Here we extend this approach to the case of systems modeled within a tight binding scheme. The grand-canonical partition function for such a system is given by
 \begin{equation}
\Theta(\mu,V,T)= \sum_{N=0}^\infty Q(N,V,T) \exp (\beta \mu N)
\label{ThetamuVT}   
\end{equation}
where $\beta=1/k_BT$, $N$ the number of atoms, $\mu$ the chemical potential of atoms and $Q(N,V,T)$ is the canonical partition function given by
 \begin{equation}
Q(N,V,T)= {V^{N} \over { N! \Lambda^{3N} } }  \int \exp\left(-\beta U(\mathbf{\Gamma})\right)d{\mathbf {\Gamma}}
\label{Q_NVT}   
\end{equation}
where $\mathbf {\Gamma}$ denotes a specific configuration of the system and $\Lambda$ is the De Broglie wavelength.

To perform an accurate sampling of the grand-canonical ensemble, it is necessary to implement a very efficient scheme for the insertion/deletion of atoms. For this purpose, we have developed an approach based on the expanded grand-canonical ensemble approach~\cite{Lyubartsev,expanded,Paul,Escobedo,Abreu,Singh,MV1,Escobedo2,MV2,Shi,Jason,Aaron,Erica,Andrew}, which consists in dividing the insertion/deletion of a full atoms into $M$ stages. Recent work has shown that the implementation of efficient schemes for the insertions/deletion steps, e.g. the expanded ensemble approach within transition matrix Monte Carlo methods~\cite{Rane} or the continuous fraction component methods~\cite{Yee,Sikora}, greatly improves the accuracy of the simulation results. Here, the combination of a Wang-Landau sampling with the expanded grand-canonical approach yields much more accurate results for the thermodynamic properties in the low temperature-high density regime, most notably for the chemical potential  \cite{PartI,Camp}. In this approach, throughout the simulation, the system contains $N$ full atoms and a fractional atom at stage $l$ (with $0 \le l \le M-1$). The coupling (or interaction) of the fractional atom with the $N$ full atoms depends on the value of $l$ and will be discussed in detail in section II.C. A fractional atom at stage $l=0$ is considered as void and does not interact with the $N$ full atoms. If $l$ is increased and its new value exceeds $M$, the fractional atom becomes a full atom and a new fractional atom at stage $l-M$ is created, leading to a system which now has $N+1$ full atoms and a new fractional atom at stage $l-M$. Similarly, if $l$ is decreased and its new value is less than $0$, the fractional atom is deleted and a randomly chosen full atom becomes a new fractional atom at stage $l+M$ is created, leading to a system which now has $N-1$ full atoms and a new fractional atom at stage $l+M$. For this system, we define a simplified expanded grand-canonical (SEGC) partition function~\cite{PartI,PartII,PartIII} as
 \begin{equation}
\Theta_{SEGC}(\mu,V,T)= \sum_{N=0}^\infty \sum_{l=0}^{M-1} Q(N,V,T,l) \exp (\beta \mu N) \\
\label{SEGC}   
\end{equation}
in which $Q(N,V,T,l)$ is the canonical partition function for a system of $N$ full atoms and a fractional atom at stage $l>0$, given by
 \begin{equation}
Q(N,V,T,l)= {V^N \over { N! \Lambda^{3N} \Lambda_ {l}^3 }  \int \exp\left(-\beta U({\mathbf {\Gamma}})\right) d{\mathbf {\Gamma}}}
\label{Q_NVTfrac}  
\end{equation}
The SEGC grand-canonical partition function differs from the conventional expanded grand-canonical partition, since it does not require the use of a weighting function (usually optimized numerically for given sets of ($N,l$) value~\cite{Trebst,Escobedo}). As discussed in previous work~\cite{PartI,PartII,PartIII}, the fact that we use a Wang-Landau sampling scheme ensures a uniform sampling of all possible $(N,l)$ values, thereby alleviating the need for a weighting function. Finally, the mass of a fractional atom (for any value of $l$ other than $0$) is chosen to be the same as that of a full atom, so that the De Broglie wavelength of the fractional atom is the same as for a full atom ($\Lambda_{l}=\Lambda$). 

\subsection{Expanded Wang-Landau sampling.}

The Wang-Landau sampling relies on an iterative evaluation of the biased distribution, $p_{bias}$~\cite{Wang1,Wang2,Shell,Shell2,Shell3,dePablo,Rampf,Camp,HMC-NPT,Copper,MolSim,MolPhys,Malakis,Hirst,KennethI,KennethII,Tsvetan}. In the context of simulations carried out in the Monte Carlo framework, we have the following Metropolis criterion for a move from an old state ($\mathbf {\Gamma_o},N_o,l_o$) to a new state ($\mathbf {\Gamma_n},N_n,l_n$)
 \begin{equation}
acc(o \to n)=min\left[ 1, {p_{bias}(\mathbf {\Gamma_n}, N_n, l_n) \over p_{bias}(\mathbf {\Gamma_o}, N_o, l_o)} \right]
\label{Metro}   
\end{equation}

In the case of a single-component system in the SEGC ensemble, the joint Boltzmann distribution $p(\mathbf {\Gamma},N,l)$ is defined as
 \begin{equation}
{p(\mathbf {\Gamma}, N, l)}={V^{N+1} \exp\left( - \beta \left[ U(\Gamma) - \mu N\right]  \right) \over { N! \Lambda^{3(N+1)}} \Theta_{SEGC}(\mu,V,T)}\\
\label{joint}   
\end{equation}
Eq.~\ref{joint} is written above for $l>0$. For a void fractional particle, $l=0$, the $(N+1)$ terms are replaced by $N$.

The number distribution $p(N,l)$ can be calculated from Eq.~\ref{joint} as
\begin{equation}
p(N, l) = \int {p(\mathbf {\Gamma}, N, l)} d\Gamma = {Q(N,V,T,l)  \exp (\beta \mu N)  \over \Theta_{SEGC}(\mu,V,T)}\\
\label{distrib}   
\end{equation}

Finally, ${p_{bias}(\mathbf {\Gamma}, N, l)}=p(\mathbf {\Gamma}, N, l) / p(N, l)$ is given by
\begin{equation}
{p_{bias}(\mathbf {\Gamma}, N, l)} = { V^{N+1}  \exp\left( - \beta \left [U(\Gamma) -\mu N \right] \right) \over { N! \Lambda^{3(N+1)} Q(N,V,T,l)}}\\
\label{pbias}   
\end{equation}

\subsection{Extension to tight-binding schemes}

In tight-binding schemes~\cite{Chadi,Sankey,DFTB,Harrison,Cohen,Menon,Elstner}, the energy includes an electronic part (calculated as the sum of single-electron energy eigenvalues of the Schr\"{o}dinger equation) and a phenomenological short-ranged repulsive part (corresponding to the repulsion between the atomic core electrons and nuclei). It is given by
\begin{equation}
\begin{array}{lll}
U & = &U_{TB}+U_{R}\\
 & = & \sum_n 2 \left< \Psi_n | H_{TB} | \Psi_n \right> + U_{R}\\
\label{TB}
\end{array}
\end{equation}
In Eq.~\ref{TB}, the repulsive energy $U_{R}$ is function of the position of atoms only, and is assumed to be independent from their electronic states.  $U_{TB}$ denotes the electronic energy, obtained from the lowest eigenvalues of the tight-bing hamiltonian $H_{TB}$ (the factor of $2$ accounting for spin). The matrix elements of the Hamiltonian $H_{TB}$ are obtained using the Slater-Koster formalism~\cite{SK} within the context of the two-center approximation of the tight-binding theory, with bonding occurring as a result of the coupling between pairs of neighboring atoms. In the case of silicon, a basis set of 4 atomic orbitals $(s,p_x,p_y,p_z)$ is assigned to each atom, and these orbitals overlap with the orbitals of neighboring atoms. For a system of $N$ atoms, the $H_{TB}$ matrix has a $4N \times 4N$ dimension and consists of $4 \times 4$ blocks - one block per atomic pair $(i,j)$. Diagonal blocks $(i=j)$ are diagonal themselves, with matrix elements along the diagonal taken to be equal to the on-site energy for the $s$ and $p$ orbitals. Off-diagonal blocks $(i \ne j)$ contain the hopping matrix elements, calculated from the distance-dependent tight-binding overlaps $h_a (r_{ij})$ (with $a$ denoting either $ss\sigma$, $sp\sigma$, $pp\sigma$ and $pp\pi$)~\cite{Goodwin,Kwon}.

Let us now consider a system containing $N$ full atoms and a fractional atom at stage $l$ $(l \ne 0)$. We now define the coupling between the fractional atom and the $N$ full atoms. The repulsive energy between a full atom and a fractional atom is rescaled with a coupling parameter $\xi_l=l/M$ in the same way as in previous work~\cite{PartI,PartII} (potential parameters with the dimension of an energy are scaled by $\xi_l^{1/3}$ and potential parameters with the dimension of a distance are scaled by $\xi_l^{1/4}$).  The electronic energy for a system of $N$ full atoms and a fractional atom is calculated as follows. The tight-binding matrix $H_{TB}$ for such a system is now chosen as a matrix of $4(N+1) \times 4(N+1)$ dimension, with the 4 additional dimensions (beyond $4N$) corresponding to the fractional atom, labeled as the $(N+1)^{th}$ particle in the system. The diagonal block for the fractional particle is given by
\begin{equation}
\left(
\begin{array}{cccc}
\epsilon_s & 0 & 0 & 0\\
0 & \epsilon_p & 0 & 0\\
0 & 0 & \epsilon_p & 0\\
0 & 0 & 0 & \epsilon_p\\
\end{array}
\right)
\label{HTBon}
\end{equation}
where $\epsilon_s$ and $\epsilon_p$ are the on-site energies for the $s$ and $p$ orbitals of the fractional atom (chosen to be the same as that of a full atom). The off-diagonal blocks for the atomic pairs involving a full atom $i$ and the fractional atom $N+1$ are given by
\begin{equation}
\left(
\begin{array}{cccc}
h^f_{ss\sigma} & d_x h^f_{sp\sigma} & d_y h^f_{sp\sigma} & d_z h^f_{sp\sigma}\\
- d_x h^f_{sp\sigma} & d_x^2 h^f_{pp\sigma}+(1-d_x^2) h^f_{pp\pi} & d_x d_y (h^f_{sp\sigma}-h^f_{sp\pi}) & d_x d_z (h^f_{sp\sigma}-h^f_{sp\pi})\\
- d_y h^f_{sp\sigma} & d_y d_x (h^f_{sp\sigma}-h^f_{sp\pi}) & d_y^2 h^f_{pp\sigma}+(1-d_y^2) h^f_{pp\pi} & d_y d_z (h^f_{sp\sigma}-h^f_{sp\pi})\\
- d_z h^f_{sp\sigma} & d_z d_x (h^f_{sp\sigma}-h^f_{sp\pi}) & d_z d_y (h^f_{sp\sigma}-h^f_{sp\pi}) & d_z^2 h^f_{pp\sigma}+(1-d_z^2) h^f_{pp\pi}\\
\end{array}
\right)
\label{HTBoff}
\end{equation}
where $h^f_{ss\sigma}$, $h^f_{sp\sigma}$, $h^f_{pp\sigma}$ and $h^f_{pp\sigma}$ are the hopping functions for the fractional-full interactions (see the 'Simulation Models' section for the detailed expression for the TB schemes used in this work) and $d_\alpha = (\alpha_i - \alpha_{N+1})/r_{i,N+1}$ (with $\alpha=x,y$ or $z$). Once the tight-binding matrix is defined, the electronic energy can be obtained by diagonalizing the $H_{TB}$ matrix and by taking the lowest eigenvalues of $H_{TB}$, considering that both full and fractional atoms have 4 valence electrons. Finally, we add that, instead of the direct diagonalization of the $H_{TB}$ matrix, linear-scaling methods can also be used to determine the electronic energy~\cite{Goedecker,Voter,Mauri,Ordejon,Drabold,Stechel, Colombo}.

\section{Simulation Models}

In this work, we use classical many-body force fields and tight-binding models for Si. The many-body force fields studied here are the widely used Stillinger-Weber (CMB-SW)~\cite{SW} and Tersoff (CMB-T)~\cite{Tersoff} potentials. The Stillinger-Weber potential $U_{CMB-SW}$ is defined as the sum of a two-body term and of a three-body term $u_3$. The two-body term between two atoms $i$ and $j$ is given by  
\begin{equation}
\begin{array}{lll}
u_2(r_{ij}) & = & A \epsilon (B (r_{ij} /\sigma )^{-p} - (r_{ij} /\sigma )^{-q}) \exp \left[ {\left( (r_{ij} / \sigma -a\right) ^{-1}} \right], (r_{ij}/ \sigma ) < a \\
 & = & 0, (r_{ij}/ \sigma ) \ge a \\
\end{array}
\label{u2SW}   
\end{equation}
where $\epsilon$, $\sigma$, $A$, $B$, $p$ and $a$ are potential parameters taken from previous work~\cite{SW}.

The three-body term between 3 atoms $i$, $j$ and $k$ is given by 
\begin{equation}
u_3({\mathbf r_{i}}, {\mathbf r_{j}}, {\mathbf r_{k}})=\epsilon \left[ h(r_{ij}, r_{ik}, \theta_{jik}) + h(r_{ji}, r_{jk}, \theta_{ijk}) +  h(r_{ki}, r_{kj}, \theta_{ikj}) \right]
\label{u3SW}   
\end{equation}
where the $h$ function is defined for $r<a$ as e.g. in the case of $h(r_{ij},r_{ik}, \theta_{jik})$
\begin{equation}
h(r_{ij},r_{ik}, \theta_{jik}) = \lambda \exp \left[ \gamma {\left( r_{ij}/\sigma -a\right) ^{-1}} + \gamma  {\left( r_{ik}/\sigma -a\right) ^{-1}}  \right] \times \left( \cos \theta_{jik} +1/3 \right)^2
\label{hSW}   
\end{equation}
where $\theta_{jik}$ denotes the angle between vectors ${\mathbf r_{ij}}$ and ${\mathbf r_{ik}}$, subtended by vertex $i$, and where $\lambda$ and $\gamma$ are potential parameters~\cite{SW}. The interaction between a full atom and the fractional atom is defined by rescaling the two-body and three-body energy terms between full atoms. For the CMB-SW potential, we use Eqs.~(12-14) with the following parameters $\epsilon_f=\xi_l^{1/3} \epsilon$ and  $\sigma_f=\xi_l^{1/4} \sigma$ for the full-fractional interaction.

The second many-body force field used in this work is the CMB-T potential~\cite{Tersoff}, that is based on a bond-order potential description of the interactions~\cite{Porter}. In this model, the interactions between two atoms $i$ and $j$ is given by  
\begin{equation}
\begin{array}{lll}
V (r_{ij}) & = & f_c (r_{ij})  \left[ A \exp \left( - \lambda r_{ij} \right) -  B \exp \left( - \mu r_{ij} \right) b_{ij} \right]\\
f_c (r_{ij}) & = & {1 \over 2}  \left[ 1 + \cos \left( {  r_{ij} -R  \over S - R } \right) \right] \\
\end{array}
\label{uTersoff}   
\end{equation}
where $b_{ij}$ is the bond order parameter, which is a many-body term that depends on the strength of the interaction between atoms $i$ and $j$, $f_c (r_{ij})$ is a cutoff function and $A$, $B$, $\lambda$, $\mu$, $S$ and $R$ are the CMB-T potential parameters. The bond order parameter directly depends on the bond geometry according to
\begin{equation}
\begin{array}{lll}
b_{ij} & = & {\left( 1 + \beta^n \zeta_{ij}^n \right)^{-1/2n} }\\
\zeta_{ij} & = & \sum_{k \ne i,j} f_c (r_{ik}) g(\theta_{ijk})\\
g(\theta_{ijk}) & = & 1 + {c^2 \over d^2} - {c^2 \over \left[ d^2+ \left( h - \cos \theta_{ijk}  \right) \right]} \\
\end{array}
\label{MBTersoff}   
\end{equation}
where $\beta$, $c$, $d$ and $h$ are potential parameters and $\theta_{jik}$ denotes the angle between vectors ${\mathbf r_{ij}}$ and ${\mathbf r_{ik}}$. We finally define the interaction between a full atom and a fractional atom for the CMB-T potential by rescaling the repulsive and attractive terms for $V (r_{ij})$ in Eq.~\ref{uTersoff}, using the following parameters $A_f=\xi_l^{1/3} A$, $B_f=\xi_l^{1/3} B$, $\mu_f=\xi_l^{1/4} \mu$ and  $\lambda_f=\xi_l^{1/4} \lambda$ for the full-fractional interaction.

The two tight-binding schemes studied in this work are the models of Kwon (QMB-K)~\cite{Kwon} and Lenosky (QMB-L)~\cite{Lenosky}. Both are orthogonal TB models, with a minimal $(s,p)$ basis and a repulsive potential, and have been shown to be highly transferable as they model accurately the properties of crystal phases, clusters~\cite{Ghasemi} as well as the solid-liquid and solid-solid phase boundaries  of silicon~\cite{Kaczmarski}. The Kwon model uses the following short-range scaling functions for the TB matrix elements~\cite{Goodwin,Kwon} for a pair of atoms $(i,j)$
\begin{equation}
h_\alpha(r_{ij})=h_\alpha(r_0) \times {\left( r_0 \over r_{ij} \right)}^n \times \exp \left[ n \left( -  {\left( r_{ij} \over r_{c\alpha} \right)}^{n_{c\alpha}} + {\left( r_0 \over r_{c\alpha} \right)}^{n_{c\alpha}} \right) \right] 
\label{KwonTB}   
\end{equation}
where $\alpha$ corresponds to either $ss\sigma$, $sp\sigma$, $pp\sigma$ or $pp\pi$ and $r_0$, $r_{c\alpha}$, $n_{c\alpha}$ and $n$ are potential parameters taken from Kwon {\it et al.}~\cite{Kwon}.
The QMB-K repulsive energy is calculated as a sum of a functional of a repulsive pair potential~\cite{Xu} $\phi(r_{ij})$
\begin{equation}
\begin{array}{lll}
U_{R} & = & \sum_i f \left[ \sum_j \phi(r_{ij})  \right]\\
f(x) & = & C_1 x + C_2 x^2 + C_3 x^3 + C_4 x^4\\
\phi_(r_{ij}) & = & {\left( r_0 \over r_{ij} \right)}^m \times \exp \left[ m \left( -  {\left( r_{ij} \over d_{c} \right)}^{m_c} + {\left( r_0 \over d_{c} \right)}^{m_c} \right) \right] 
\end{array}
\label{KwonRep}   
\end{equation}
where $C_n$ ($n=1$, $2$, $3$ or $4$), $r_0$, $d_c$, $m$ and $m_c$ are potential parameters~\cite{Kwon}.

For the QMB-K model, we model the interaction between a full atom and a fractional atom as follows. We use the same functional form $h_{\alpha}(r_{ij})$ defined in Eq.~\ref{KwonTB} and use the following scaled parameters: $h^f_\alpha(r_0)=\xi_l^{1/3} h_\alpha(r_0)$ (with $\alpha$ corresponding to either $ss\sigma$, $sp\sigma$, $pp\sigma$ or $pp\pi$), $r^f_0=\xi_l^{1/4} r_0$ and $r_{c\alpha}=\xi_l^{1/4} r_{c\alpha}$. The electronic energy is then obtained by determining the lowest eigenvalues of the $4 (N+1) \times 4(N+1)$ matrix (for a system of $N$ full atoms + 1 fractional particle), keeping in mind that the full atoms as well as the fractional atom all have 4 valence electrons. For the repulsive part, we use the same set of equations (Eq.~\ref{KwonRep}) with parameters scaled as follows: $C^f_n=\xi_l^{1/3} C_n$ ($n=1$, $2$, $3$ or $4$), $r^f_0= \xi_l^{1/4} r_0$ and $d^f_c= \xi_l^{1/4} d_c$.

\begin{figure}
\begin{center}
\includegraphics*[width=10cm]{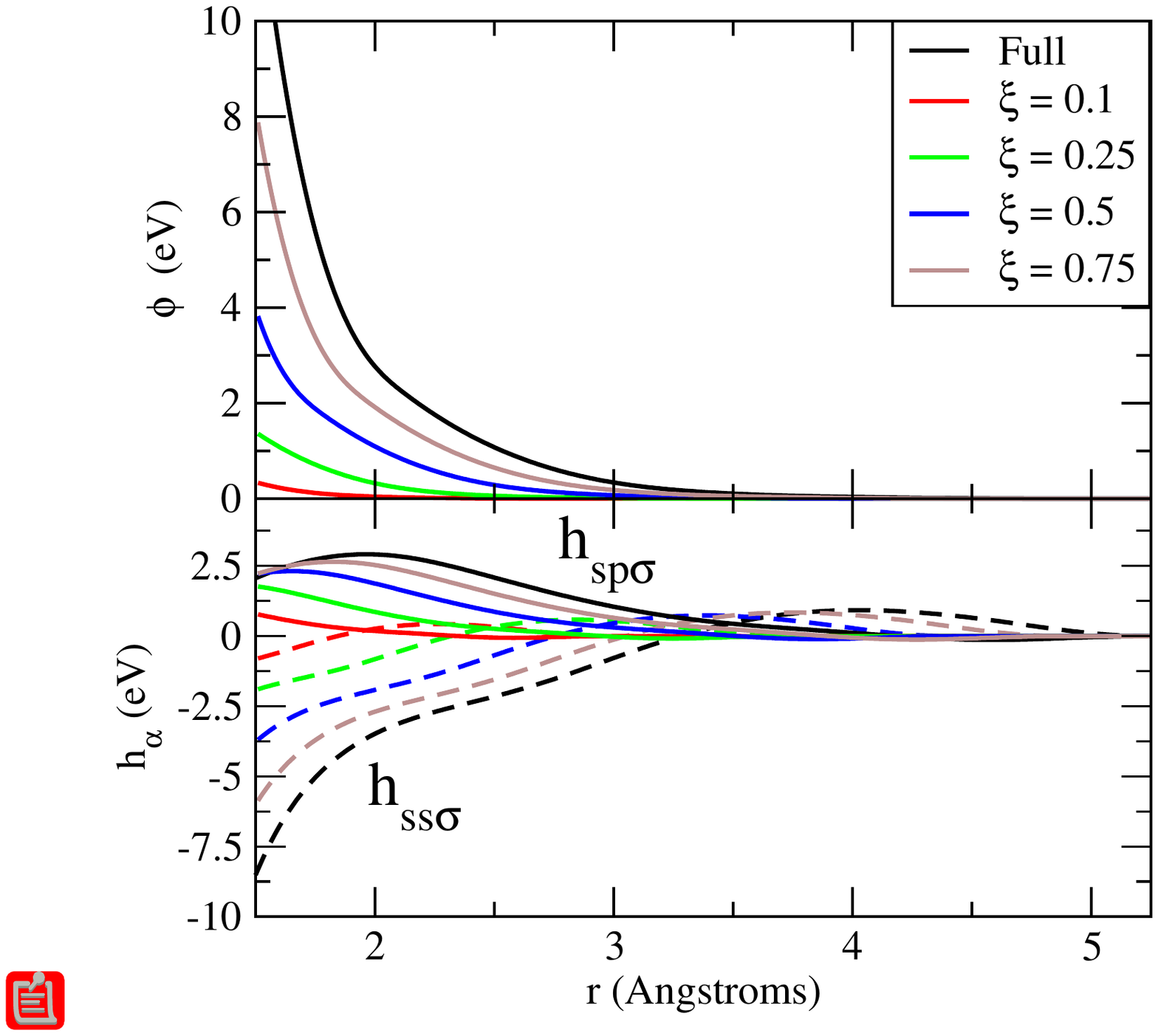}(a)
\includegraphics*[width=10cm]{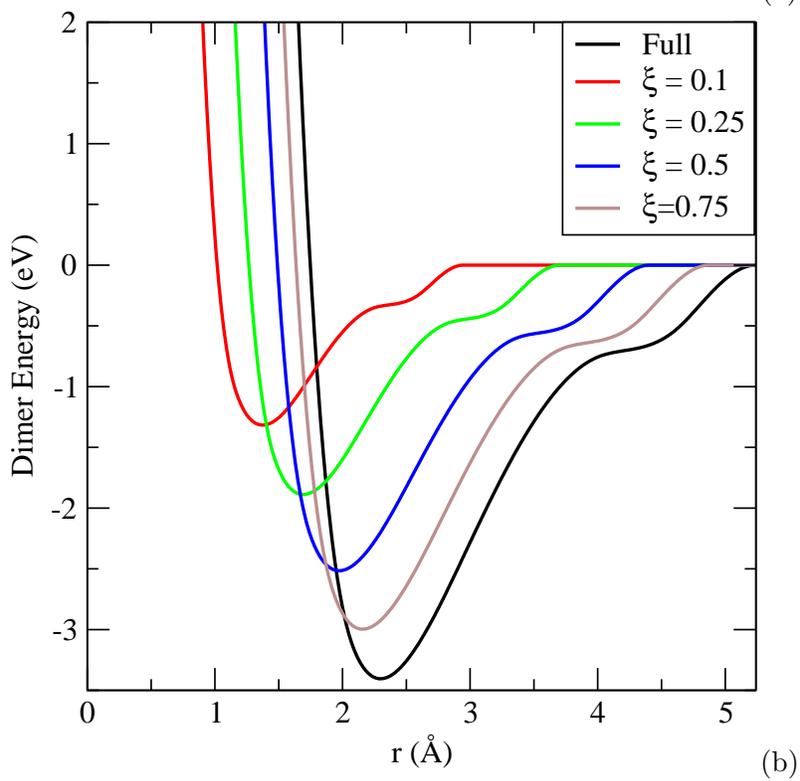}(b)
\end{center}
\caption{DESGRANGES-DELHOMMELLE}
\label{Fig1}
\end{figure}

The QMB-L model~\cite{Lenosky} is also an orthogonal TB model, that uses cubic splines to represent the 4 functions defining the TB matrix elements as well as the pair repulsive potential. To model the interaction between a full atom and a fractional atom, we apply a rescaled version of the cubic splines defined by Lenosky~{\it et al.}. In the QMB-L model, the TB matrix elements are defined from the scaling functions $h_{\alpha}(r)=g_{\alpha}(r)/r^2$, where $r$ is the interatomic distance for a pair of atoms, $g_{\alpha}$ is a cubic spline and $\alpha$ one of the 4 possible overlaps ($\alpha={ss\sigma}$, ${sp\sigma}$, ${pp\sigma}$ or ${pp\pi}$). The scaling function for a pair including a full atom and a fractional atom is obtained by rescaling the energies by ${\xi_l}^{1/4}$ and the distances by ${\xi_l}^{1/3}$, leading to $h^f_{\alpha}(r)={\xi_l}^{1/4}g_{\alpha}(r/{\xi_l}^{1/3})/(r/{\xi_l}^{1/3})^2$, and the scaled pair potential for the full-fractional repulsion is given by: $\phi^f (r_{ij})= \xi^{1/4} \phi (r_{ij} / \xi^{1/3})$. As for the QMB-K model, the potential energy is obtained by adding the repulsive energy to the electronic energy, obtained from the lowest eigenvalues of the TB matrix. Fig.~\ref{Fig1}(a) summarizes the dependence of the full-fractional pair potential on the coupling parameter. As the coupling parameter decreases, the repulsive pair potential smoothly decreases, ensuring that the insertion of the fractional atom is facilitated. Similarly, the TB matrix elements decrease with the coupling parameter, which leads to a smooth transition from a fractional atom so a full atom as it is grown during the EWL simulations. The resulting potential energy (calculated as the sum of the repulsive and electronic contributions) for dimers, composed of a full atom and of a fractional atom, are shown in Fig.~\ref{Fig1}(b) for different values of the coupling parameters. As the coupling parameter increases, the minimum is smoothly shifted from a shorter distance (reached at $1.37$~\AA~for an energy of $1.3$~eV for $\xi=0.1$) to a larger distance (e.g. reached at $2.14$~\AA~for an energy of $3$~eV for $\xi=0.9$), until the full Lenosky dimer energy is recovered (with a minimum of $3.4$~eV reached for a distance of $2.28$~\AA, in very good agreement with the {\it ab initio} results~\cite{Sinnott}).

EWL simulations are performed on systems of up to 200 atoms within the framework of Monte Carlo (MC) simulations, with the following two types of MC moves: (i) a translation of a single, full or fractional, atom ($75\%$ of the MC steps) and (ii) a change in $(N,l)$ ($25\%$ of the MC steps). The technical details regarding the Wang-Landau scheme are the same as described in the first papers of the series~\cite{PartI,PartII,PartIII}. To study the properties of the fluid phases of Silicon under nanoscopic confinement, we use a slit pore geometry and model the interactions between the fluid atoms and the two confining walls with the well-known Steele 9-3 potential~\cite{gelb1999,Steele}. More specifically, the fluid is confined between 2 planar walls separated by a distance of $12$~\AA~along the $z$-axis. The effective interaction between the atoms of the fluid with the top wall (located at $S_z/2=6$~\AA~along the $z$-axis) is given by
\begin{equation}
\phi_{wf}^t(z)= {2 \pi \rho_w \sigma_{w-Si}^3 \epsilon_{w-Si} \over 3} \left[ {2 \over 15} \left( \sigma_{w-Si} \over S_z/2 - z\right)^9 -  \left( \sigma_{w-Si} \over S_z/2 - z\right)^3 \right]
\label{wallfluidT} 
\end{equation}
while the interaction with the bottom wall (located at $-S_z/2=-6$~\AA~along the $z$-axis) is 
\begin{equation}
\phi_{wf}^b(z)= {2 \pi \rho_w \sigma_{w-Si}^3 \epsilon_{w-Si} \over 3} \left[ {2 \over 15} \left( \sigma_{w-Si} \over S_z/2 + z\right)^9 -  \left( \sigma_{w-Si} \over S_z/2 + z\right)^3 \right]
\label{wallfluidB} 
\end{equation}
For the wall-fractional atom interaction, we simply use a rescaled version of Eqs.~\ref{wallfluidT}~and~\ref{wallfluidB} using as interaction parameters $\epsilon^{f}_{w-Si}=\xi_l^{1/3}\epsilon_{w-Si}$ and $\sigma^{f}_{w-Si}=\xi_l^{1/4}\sigma_{w-Si}$. We consider a graphite wall (using the parameters for carbon determined by Steele~\cite{Steele}, with a number density for the wall $\rho_w=0.097$~atoms/\AA$^{3}$ taken from previous work~\cite{Porcheron,Padilla}) and model the wall-fluid interactions using the Silicon parameters of Murad and Puri~\cite{Murad}, which gives $\sigma_{w-Si}=3.71$~\AA~and $\epsilon_{w-Si}/k_B=54.15$~K.

\section{Results and Discussion}

\subsection{Thermodynamics of the fluid phases of Si}

\begin{figure}
\begin{center}
\includegraphics*[width=10cm]{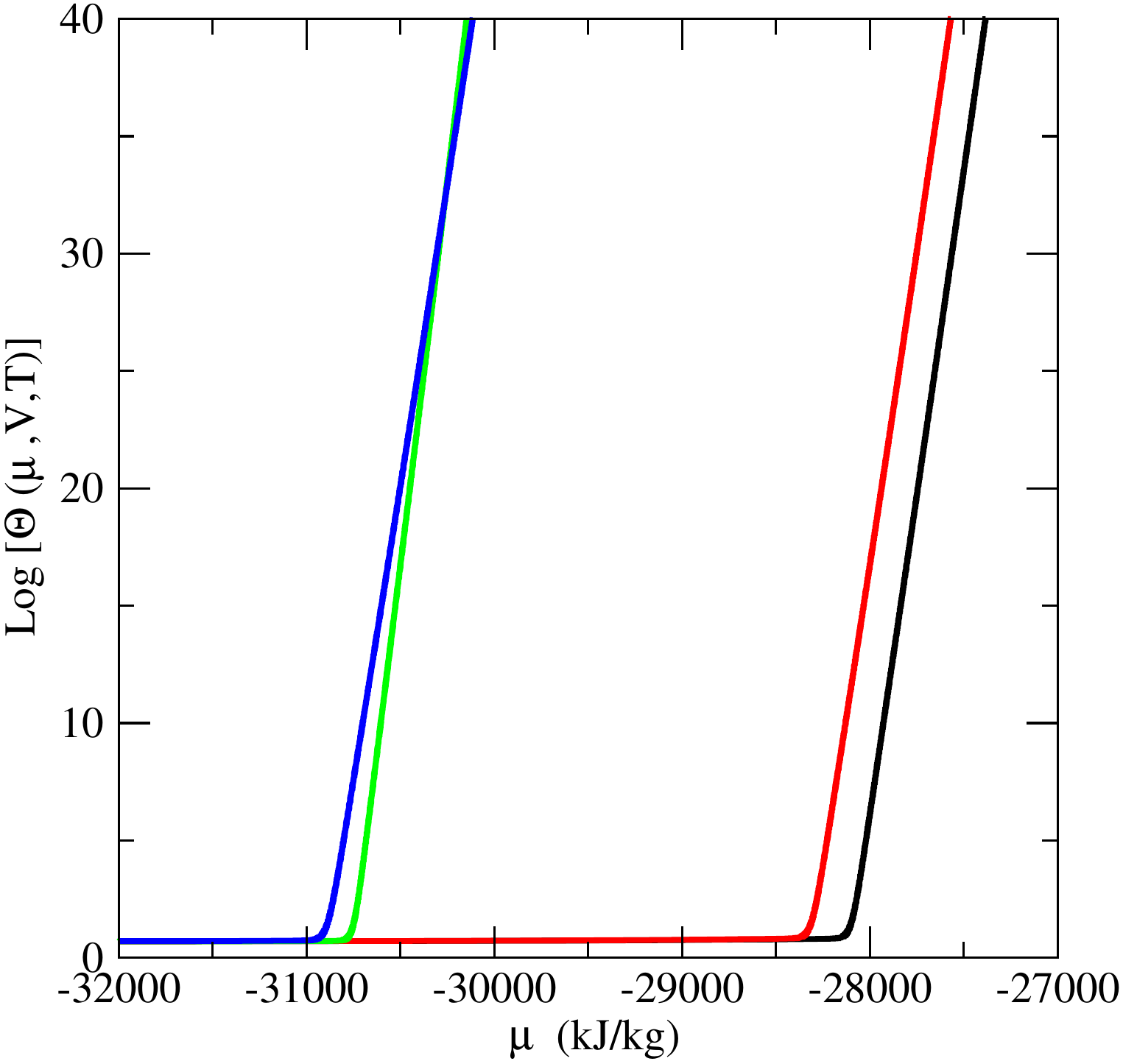}(a)
\includegraphics*[width=10cm]{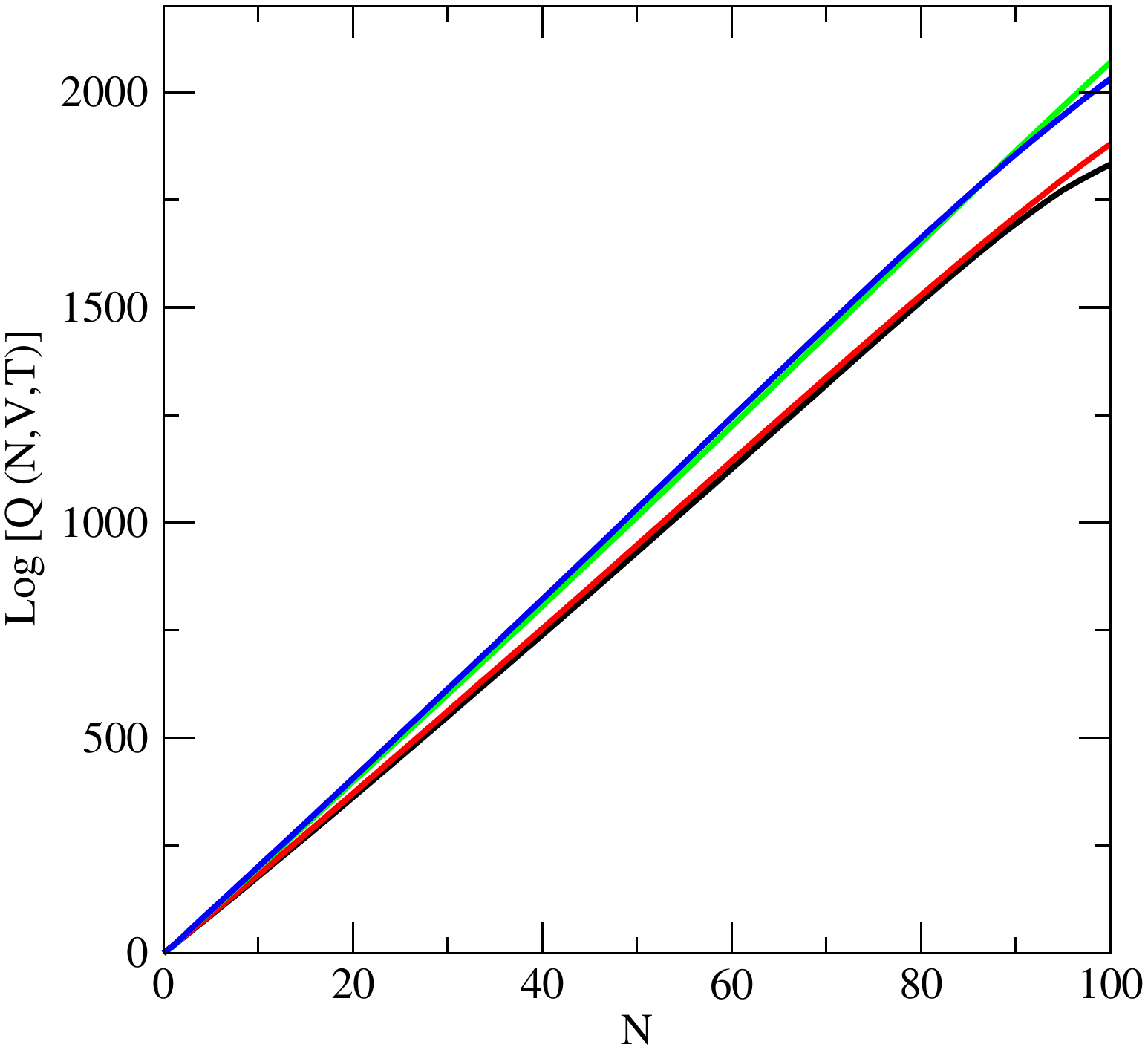}(b)
\end{center}
\caption{DESGRANGES-DELHOMMELLE}
\label{Fig2}
\end{figure}

We plot in Fig.~\ref{Fig2}(a) the results obtained, at a temperature $T=5000$~K, for the grand-canonical partition function $\Theta (\mu, V, T)$ of Si modeled with the two classical many-body force fields (CMB-SW and CMB-T) and the two quantum many-body models (QMB-K and QMB-L) considered in this work. For low chemical potentials ($\mu < -31000$~kJ/kg), the values for the partition function predicted by the different models are in excellent agreement with each other. As the chemical potential increases, Fig.~\ref{Fig2}(a) shows that the grand-canonical partition sharply increases, first for the two QMB models, and then for the two CMB models. The sudden increase in $\Theta (\mu,V,T)$ is first observed for the QMB-L model ($\mu > -30900$~kJ/kg), then shortly after for the QMB-K model ($\mu > -30750$~kJ/kg) and then for the CMB-T model ($\mu > -28350$~kJ/kg) and finally for the CMB-SW model ($\mu > -28150$~kJ/kg). This sharp increase in the partition function indicates the onset of a phase transition from a low density phase towards a phase of higher density (in our case here, the vapor-liquid transition for Silicon). This means that the predicted value for the chemical potential at coexistence will be the lowest for the QMB-L model, followed by the QMB-K model, the CMB-T model and finally the CMB-SW model. The predictions by the two QMB models for $\mu$ at coexistence are close (within $0.5$~\% of each other) and around $8$~\% below the CMB predictions. As the chemical potential further increases beyond its value at coexistence, Fig.~\ref{Fig2}(a) shows that, for all models, the grand-canonical partition function increases steadily with the chemical potential.

The grand-canonical partition function is obtained by summing up the functions $Q(N,V,T)$, with a weighting factor of $\exp[\beta \mu N]$ (see Eq.~\ref{ThetamuVT}). The onset of the vapor-liquid transition, indicated by a sharp increase in the grand-canonical partition function $\Theta (\mu, V, T)$, can also be identified on a plot of $Q(N,V,T)$ against $N$. Fig.~\ref{Fig2}(b) shows that the values for the slopes of the logarithm of $Q(N,V,T)$ are in the following order, QMB-L~$>$~QMB-K~$>$~CMB-T~$>$~CMB-SW. This means that the order observed for the slopes is the opposite of the order obtained for $\mu$ at coexistence shown in Fig.~\ref{Fig2}(a). This can be understood in terms of the number distribution $p(N)$ given by
\begin{equation}
p(N) = {Q(N,V,T) \exp{\beta \mu N} \over \Theta(\mu,V,T)}\\
\label{Equal}
\end{equation}
$p(N)$ is proportional to the weighted function $Q(N,V,T) \exp{\beta \mu N}$. At coexistence, the number distribution at low $N$ (corresponding to the vapor) and at large $N$ (corresponding to the liquid) contribute equally. This occurs when $Q(N,V,T) \exp{\beta \mu N}$ take similar values for both phases, or, equivalently, when $\log Q(N,V,T)$ exhibits a slope that compensates (i.e. is of the opposite sign) the linear term in $N$, $\beta \mu N$. This accounts for the fact that the values for the slopes for $\log Q(N,V,T)$ are in the opposite order of that obtained for $\mu$ at coexistence.

The next step consists in  using the results obtained for the partition functions to determine the thermodynamic properties of the fluid phases of Si, as predicted by the different CMB and QMB models studied here. The number distribution $p(N)$ can be used to determine numerically the value of the chemical potential at the vapor-liquid coexistence for each model. At coexistence, the two phases have equal probabilities. Noting $\Pi_l$ the probability associated with the liquid and $\Pi_v$ the probability associated with the vapor, writing the condition that $\Pi_l=\Pi_v$ leads to
\begin{equation}
\sum_{N<N_b} p(N)=\sum_{N>N_b} p(N)\\
\label{Equal2}
\end{equation}
with $N_b$ denoting the boundary value (the value of $N$ for which $p(N)$ reaches a minimum between the two peaks corresponding to the liquid and vapor phases).

\begin{figure}
\begin{center}
\includegraphics*[width=10cm]{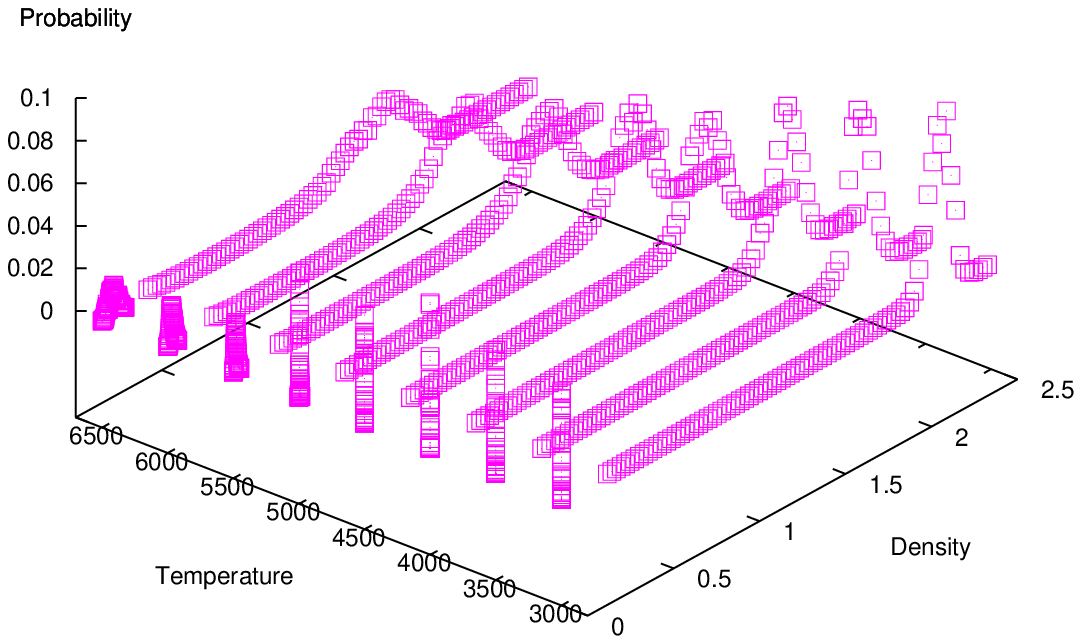}(a)
\includegraphics*[width=10cm]{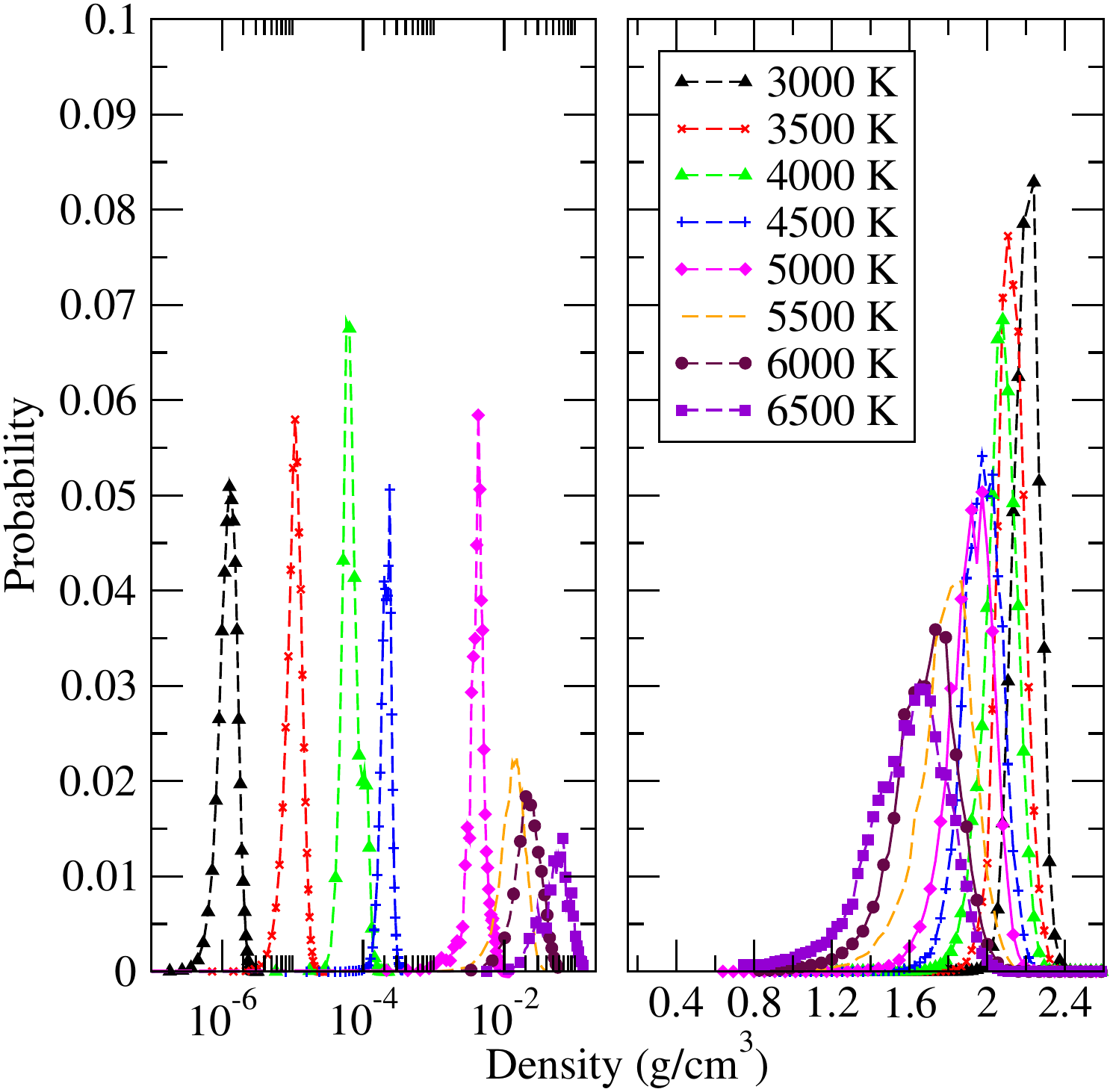}(b)
\end{center}
\caption{DESGRANGES-DELHOMMELLE}
\label{Fig3}
\end{figure}

We show in Fig.~\ref{Fig3} the results obtained for the density distribution $p(\rho=N/V)$ at coexistence, in the case of the QMB-L model. For each temperature, $p(\rho)$ exhibits two peaks, associated with either the vapor phase for low densities or the liquid phase for high densities. Graphing the number distribution against the density and the temperature allows us to obtain the 3-D plot shown in Fig.~\ref{Fig3}(a). This 3-D plot is especially interesting since it provides a 3-D sketch of the phase diagram, drawn on the basis of the phase probabilities, with the two series of peaks underlying the phase envelope. To give a better account of the variation, as a function of temperature, of the densities of the two phases at coexistence, we plot in Fig.~\ref{Fig3} the probabilities obtained for each phase, either on a logarithmic scale for the vapor (see the left-hand-side of Fig.~\ref{Fig3}(b)) or on a linear scale for the liquid (see the right-hand-side of Fig.~\ref{Fig3}(b)). These plots show the expected behavior for the densities of the two phases at coexistence, with a shift of the maximum probability for the vapor towards larger values for the density as the temperature increases (with a density increase of $5$ orders of magnitude as the temperature goes from $3000$~K to $6500$~K) and a shift of the maximum probability for the liquid towards lower densities as the temperature increases (with a decrease in density by 27~\% as the temperature goes from $3000$~K to $6500$~K) .

\begin{figure}
\begin{center}
\includegraphics*[width=10cm]{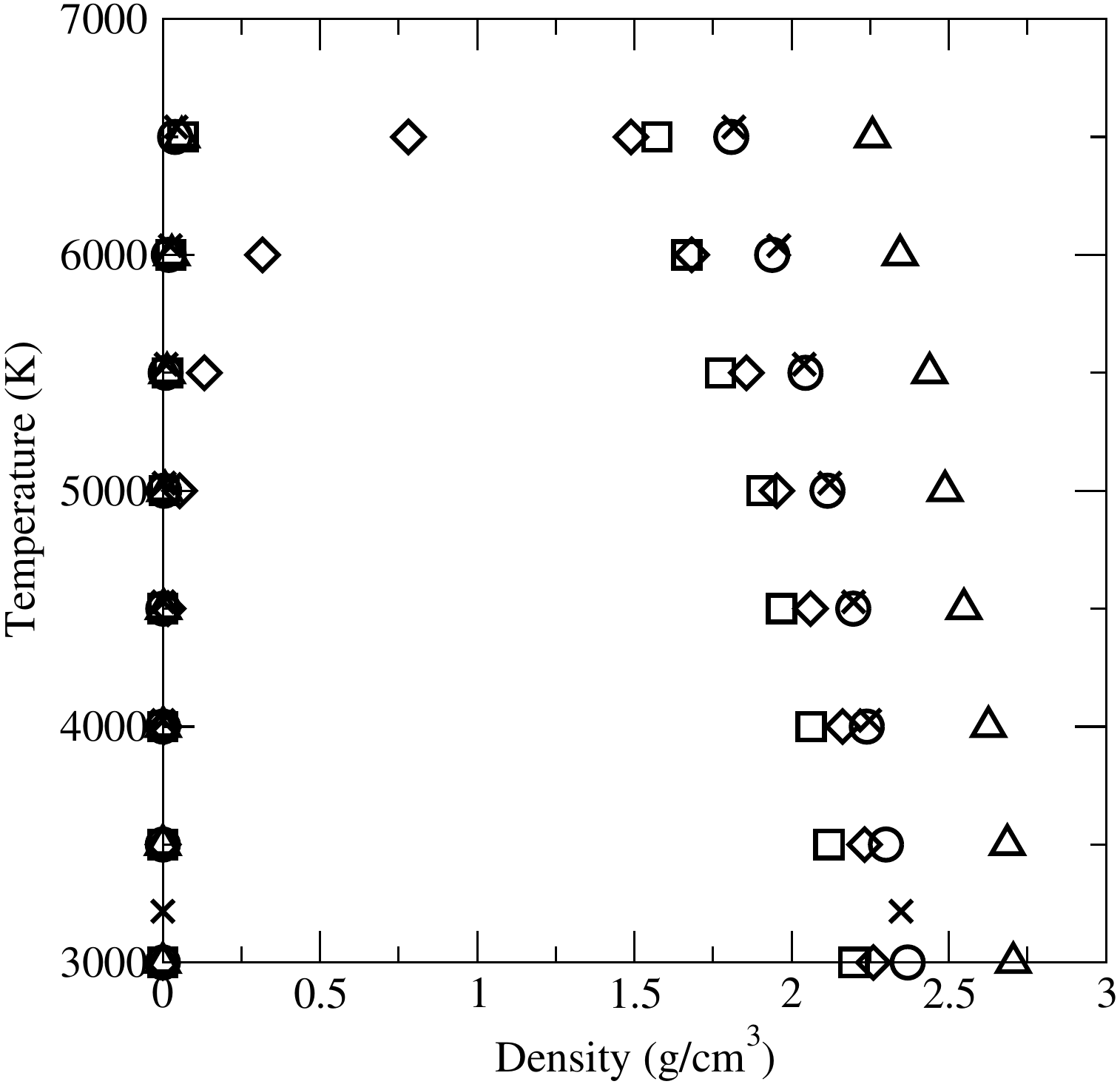}(a)
\includegraphics*[width=10cm]{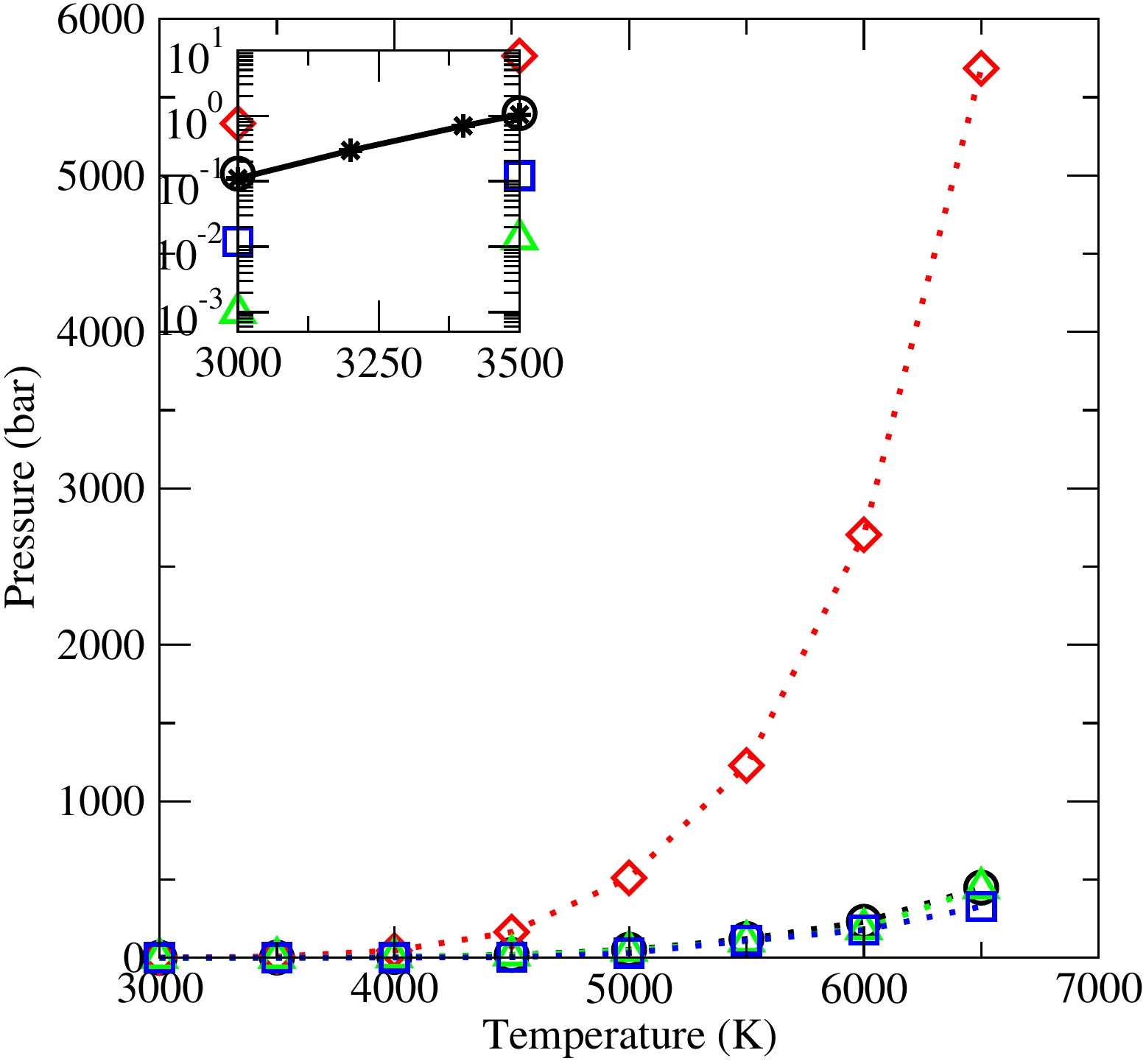}(b)
\end{center}
\caption{DESGRANGES-DELHOMMELLE}
\label{Fig4}
\end{figure}

The vapor-liquid equilibria for each model can be plotted in the temperature-density plane by reporting the densities at coexistence for the liquid, $<\rho_l>$, and for the vapor, $<\rho_v>$, obtained from
\begin{equation}
\begin{array}{lll}
<\rho_l> & = & {\sum_{N>N_b} (N/V)p(N) \over {\sum_{N>N_b} p(N)}}\\
<\rho_v> & = & {\sum_{N<N_b} (N/V)p(N) \over {\sum_{N<N_b} p(N)}}\\
\end{array}
\label{probabilities}   
\end{equation}
The densities at coexistence are shown in Fig.~\ref{Fig4}(a) for the CMB and QMB models for temperatures ranging from $3000$~K and $6500$~K. For comparison purposes, we also include the results~\cite{Honda} from prior work, obtained using the CMB-SW model and the Gibbs Ensemble Monte Carlo (GEMC) algorithm~\cite{Panagiotopoulos1,Panagiotopoulos2}. The GEMC results for the CMB-SW are found to be in excellent agreement with the EWL results obtained in this work. While the chemical potentials at coexistence were found to be similar within the same class of model (either CMB or QMB - see Fig.~\ref{Fig2}), a similar type of classification does not strictly apply to the results for the densities. Fig.~\ref{Fig4}(a) shows that the EWL densities  are very sensitive to the model parameters. For instance, the density of the liquid phase at coexistence is found to be the highest for the QMB-K model (close to e.g. $2.7~g/cm^3$ at $3000$~K) and the lowest for the QMB-L model (about $2.2~g/cm^3)$. On the other hand, the density of the vapor is found to be lower for the QMB models (with the QMB-K density being roughly an order of magnitude less than for the QMB-L model over the temperature range considered here) than for the CMB models (with the CMB-SW density being an order of magnitude less than for the CMB-T model over the temperature range). Overall, on the basis of the VLE densities, a critical comparison of the various models leads to the following conclusion. The two models exhibiting extreme behaviors are observed for the QMB-K model (highest liquid density-lowest vapor density) and for the CMB-T model (second lowest liquid density-highest vapor density), while intermediate behaviors are observed by the CMB-SW and QMB-L model. 

\begin{figure}
\begin{center}
\includegraphics*[width=10cm]{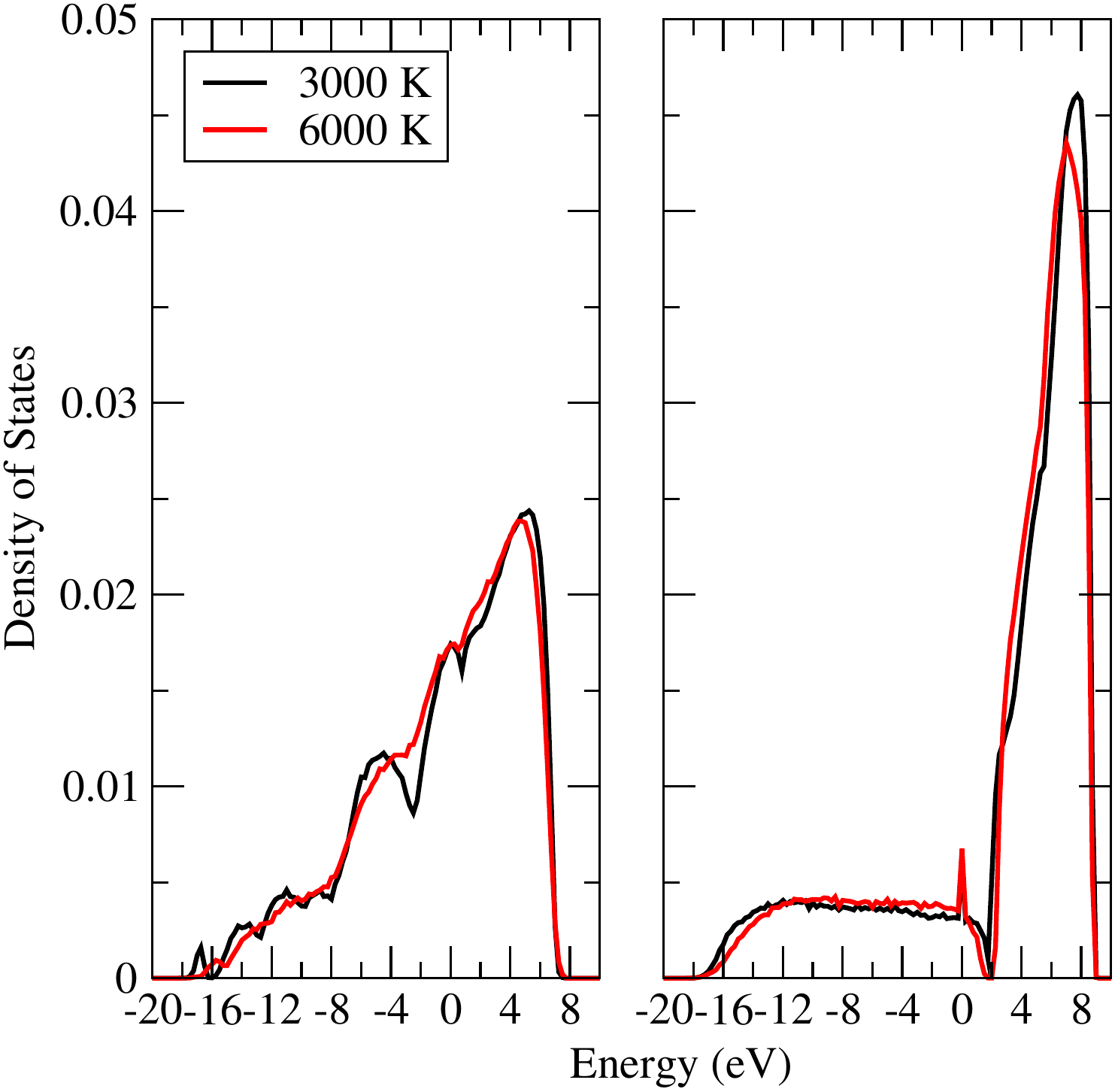}
\end{center}
\caption{DESGRANGES-DELHOMMELLE}
\label{Fig5}
\end{figure}

While experimental data are not available for the densities at coexistence, the experimental data for the vapor pressure~\cite{Desai} provide a way to assess the performance of the force fields to model the fluid phases of Si. In EWL simulations, the pressure at coexistence can be determined from $\Theta(\mu,V,T)$ through the following equation (using the value at coexistence for $\mu$)
\begin{equation}
P = {k_B T \ln \Theta(\mu,V,T) \over V}\\
\label{Pressure}   
\end{equation}
The EWL results for the pressure at coexistence are shown in Fig.~\ref{Fig4}(b) for the 4 models and compared to the available experimental data~\cite{Desai}. The results confirm the conclusions drawn from the results for the densities at coexistence, with the results for QMB-K and CMB-T appearing to be outliers. The QMB-K pressure underestimates by almost 2 orders of magnitude the experimental data at low temperatures, which is consistent with the very low density predicted for the vapor. This behavior could be attributed to the definition of the QMB-K model, which gives an incorrect energy for isolated atoms~\cite{Lenosky}. Such cases occur frequently and thus become of great significance at low temperatures, when the vapor density becomes very low. The CMB-T pressure consistently overestimates the experimental data as well as the pressures predicted by all other models, especially at high temperatures where the vapor density predicted by the CMB-T model appears to be too large. In line with the results for the densities, the CMB-T and QMB-K results bracket the predictions by the CMB-SW model (which are remarkably close to the experimental data) and by the QMB-L model. The QMB models also provide access to additional information, that cannot be obtain when classical many-body force fields are used. These include the electronic density of state plotted in Fig.~\ref{Fig5} for the two QMB models studied in this work. Fig.~\ref{Fig5} shows the electronic density of states for the liquid, along the coexistence line, at $T=3000$~K and $T=6000$~K. The QMB-K model predicts that the electronic density of states retains the same qualitative features as temperature increases, consistently with the very moderate change in liquid density with the temperature (the liquid density at coexistence decreases by only 13.3~\% for the QMB-K model as the temperature increases from $3000$~K to $6000$~K). On the other hand, increasing the temperature results in a qualitatively different electronic density of states for the QMB-L model, with the formation of a gap around $1.9$~eV at high temperature (see the right panel of Fig~\ref{Fig5}) as a result of the larger change in liquid density observed for this model (the liquid density decreases by 24.2~\% for the QMB-L model. compared to 13.3\% for the QMB-K model, over the same temperature interval).

In the rest of the paper, we focus on the results obtained for the CMB-SW and QMB-L models, since these have been shown to provide predictions that are closest to the experiment for the vapor-liquid coexistence. We compare the results obtained for compressed liquid for the CMB-SW and QMB-L model at $T=5000~K$ and for pressures ranging from $0.2$~GPa to $1$~GPa. Comparing the two sets of results, we find that the predictions for the properties of compressed liquids exhibit similar deviations to those observed for the vapor-liquid phase diagram. We find a shift of the order of $2$ to $3 \times 10^3~kJ/kg$ for the Gibbs free energy (in line with the results for the chemical potentials at coexistence) and larger densities for the CMB-SW model, by up to about 9~\% (consistently with the findings for the liquid densities at coexistence). Both models are found to capture the essential features of the thermodynamic response of Si to the increase in pressure. In particular, the effect of pressure on the density is of the same order for both models. The dependence upon pressure of the thermodynamic properties is also found to be very similar for both models, with a similar increase in Gibbs free energy over the pressure range (about $400~kJ/kg$ for both models) and in the enthalpy (close to $250~kJ/kg$ for both models) and a similar decrease in entropy over the pressure range ($0.02~kJ/kg$ for both models).

\subsection{Thermodynamics of nanoconfined Si}

\begin{figure}
\begin{center}
\includegraphics*[width=10cm]{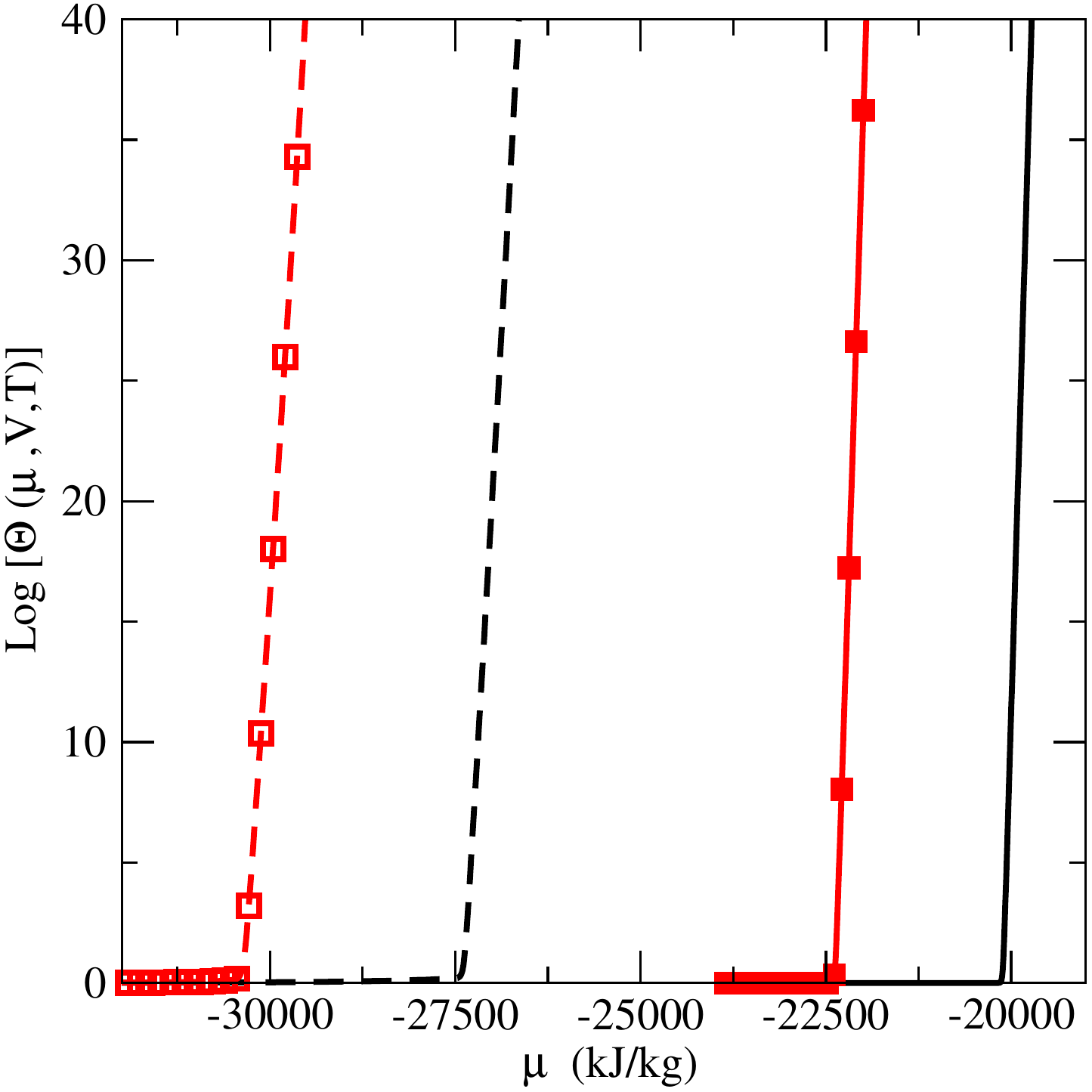}(a)
\includegraphics*[width=10cm]{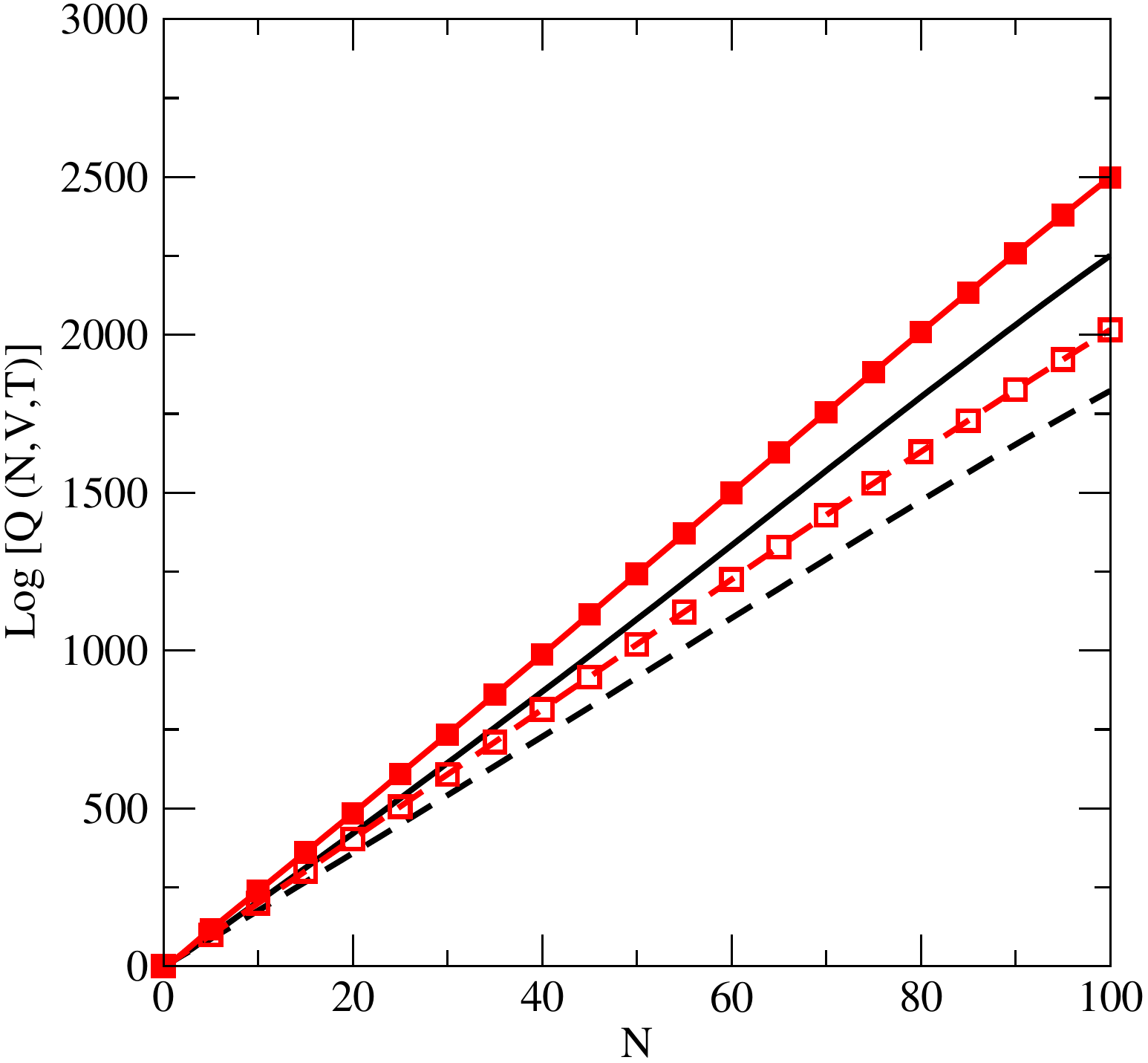}(b)
\end{center}
\caption{DESGRANGES-DELHOMMELLE}
\label{Fig6}
\end{figure}

We now apply the EWL method to determine the partition functions for nanoconfined fluid phases of Si. We begin our analysis with the results obtained for the grand-canonical partition function $\Theta (\mu,V,T)$. Fig.~\ref{Fig6}(a) shows the partition functions obtained for nanoconfined Si with the two models for $T=3000$~K and $T=5000$~K. The grand-canonical partition functions are in very good agreement for the two models at low chemical potentials (e.g. for $\mu<-22350$~kJ/kg at $T=3000$~K). As the chemical potential increases, the partition functions for the models start to depart strongly from each other. The QMB-L partition function exhibits a steep increase when $\mu$ exceeds $-22350$~kJ/kg, while the CMB-SW partition function increases rapidly once $\mu$ becomes greater than $-20100$~kJ/kg. A similar behavior is observed at higher temperatures, as shown for $T=5000$~K on the left of Fig.~\ref{Fig6}(a), with the QMB-L grand-canonical partition function increasing steeply for a chemical potential about $2900$~kJ/kg less than for the CMB-SW partition function. The steep increase in the partition function is associated with a transition from a nanoconfined phase of low density to a nanoconfined phase of high density. Comparing the results for nanoconfined Si to those obtained for the bulk, we find that, for a given model (either CMB-SW or QMB-L), confining Si results in a shift of the transition, from the low density phase to the high density phase, in $\mu$ by about $650$~kJ/kg at $T=5000$~K. Examining the results for $\log Q(N,V,T)$ provides a direct way to locate the chemical potentials for which the phase transition takes place. At coexistence, the chemical potential is such that the linear term $\beta \mu N$ takes the opposite value of the slope of $\log Q(N,V,T)$. For a given temperature (fixed $\beta$), Fig~\ref{Fig6}(b) shows that the slope of $\log Q$ is greater for the QMB-L model than for the CMB-SW model. This means that the transition from the low-density nanoconfined phase to the high-density nanoconfined phase occurs at lower $\mu$ for QMB-L than for CMB-SW, consistently with the delay in $\mu$ for the steep increase of $\Theta_{CMB-SW}$ when compared to $\Theta_{QMB-L}$ (see Fig.~\ref{Fig6}(a)).

\begin{figure}
\begin{center}
\includegraphics*[width=12cm]{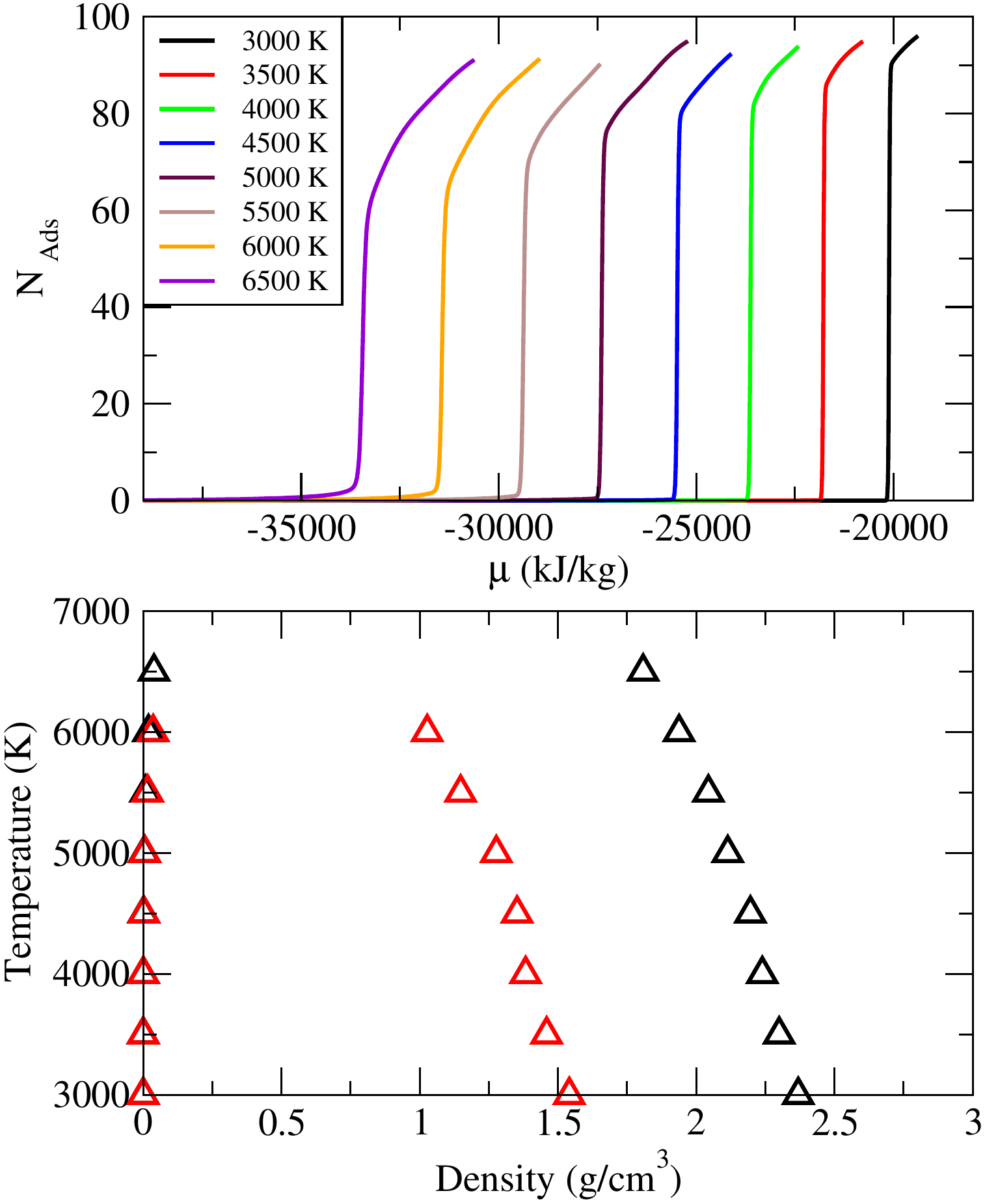}
\end{center}
\caption{DESGRANGES-DELHOMMELLE}
\label{Fig7}
\end{figure}

\begin{figure}
\begin{center}
\includegraphics*[width=12cm]{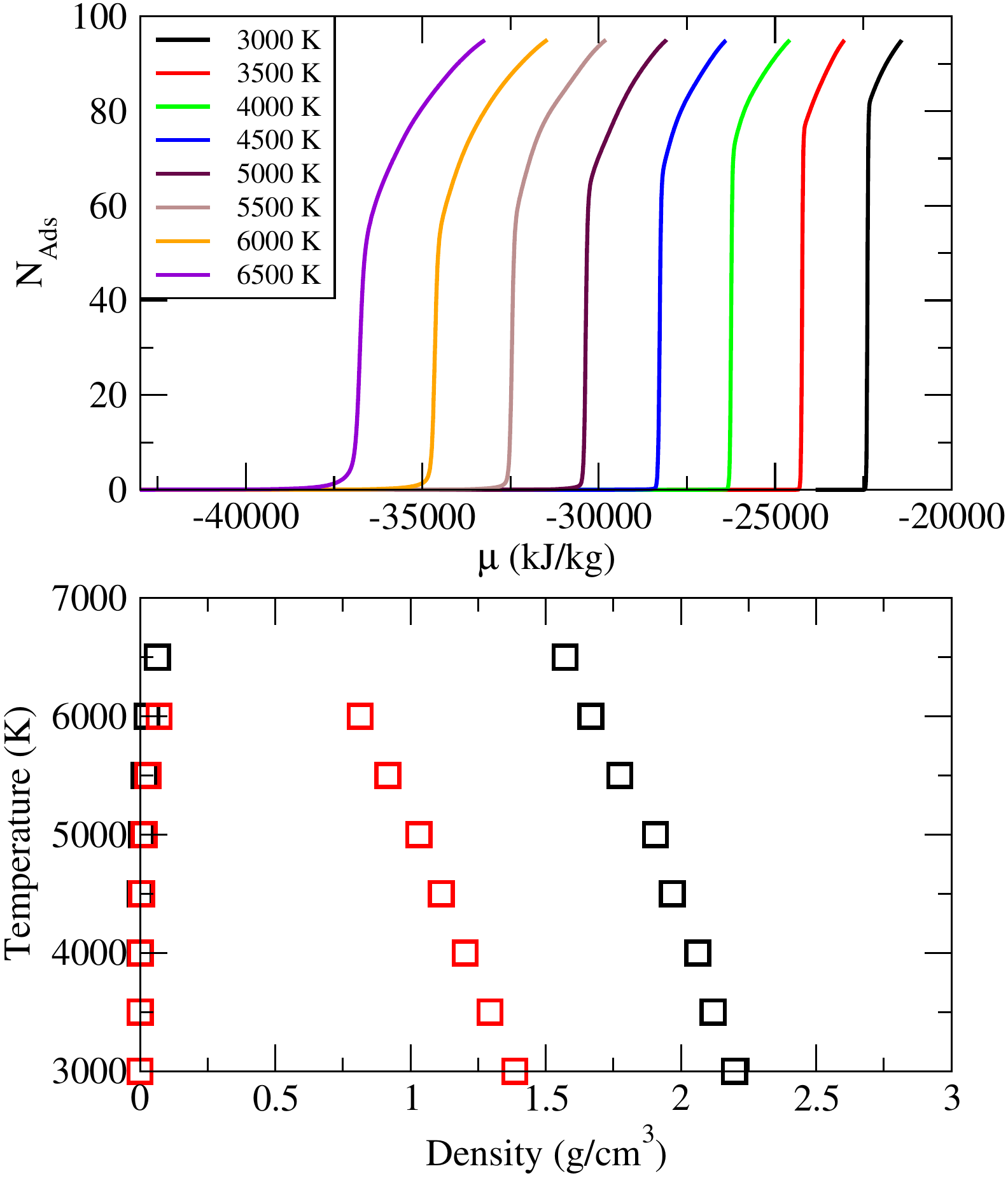}
\end{center}
\caption{DESGRANGES-DELHOMMELLE}
\label{Fig8}
\end{figure}

We now turn to the thermodynamics of nanoconfined fluid phases of Si. Starting with the predictions from the CMB-SW model, we plot in Fig.~\ref{Fig7}(a) the adsorption isotherms for Si for temperatures ranging from $3000$~K to $6500$~K. The adsorption isotherms exhibit a step, corresponding to the transition from the low-density to the high-density phase, for a value of the chemical potential that coincides with the chemical potential marking the steep increase in $\log \Theta$ in Fig.~\ref{Fig6}(a) (e.g. slightly below $-20000$~kJ/kg at $T=3000$~K). As the temperature increases, the adsorption step is shifted towards the lower end of the range for the chemical potential, in agreement with the behavior of $\log \Theta$ as a function of $T$ in Fig.~\ref{Fig6}(a). To analyze further the EWL results, we determine the phase diagram for nanoconfined Si by evaluating the properties of the the two coexisting phases from the EWL data. For this purpose, we calculate the void volume~\cite{Myers,Jason,Aaron} $V_{void}$, i.e. the volume of the nanopore accessible to the fluid and find a value of $V_{void}=0.973V_{pore}$ ($V_{pore}$ being the total volume of the nanopore considered here). The density of the two nanoconfined phases is then obtained from the number distribution $p(N)$ through Eq.~\ref{probabilities}. The phase diagram so obtained for nanoconfined Si is plotted in Fig.~\ref{Fig7}(b) and compared to the EWL results for the bulk. Nanoconfinement strongly impacts the phase coexistence. For a given temperature, subjecting Si to a nanoscopic confinement leads to an increase in the density of the vapor at coexistence (e.g. by $93$~\% at $T=6000$~K) and to a decrease in the density of the liquid at coexistence (by $47$~\% at $T=6000$~K). The adsorption isotherms for Si predicted by the QMB-L model are plotted in Fig.~\ref{Fig8}(a). Fig.~\ref{Fig8}(a) shows that the adsorption step occurs concomitantly with the steep increase in $\log \Theta$ seen in Fig.~\ref{Fig6}(a). This further confirms the connection between the steep increase in the partition function and the phase transition taking place in the nanoconfined fluid. Moreover, as temperature increases, the adsorption step occurs for lower values of the chemical potential, consistently with the findings for the CMB-SW model. The phase diagram for nanoconfined Si predicted by the QMB-L model is compared to the vapor-liquid equilibrium for the bulk in Fig.~\ref{Fig8}(b). As for the CMB-SW model, the QMB-L model predicts that the coexistence curve for nanoconfined Si is located inside the phase envelope of the bulk. Nanoconfinement results in sharp changes in the densities of the two coexisting phases, with an increase in the density of the vapor at coexistence (e.g. by $173$~\% at $T=6000$~K) and to a decrease in the density of the liquid at coexistence (by $51$~\% at $T=6000$~K).

\begin{figure}
\begin{center}
\includegraphics*[width=16cm]{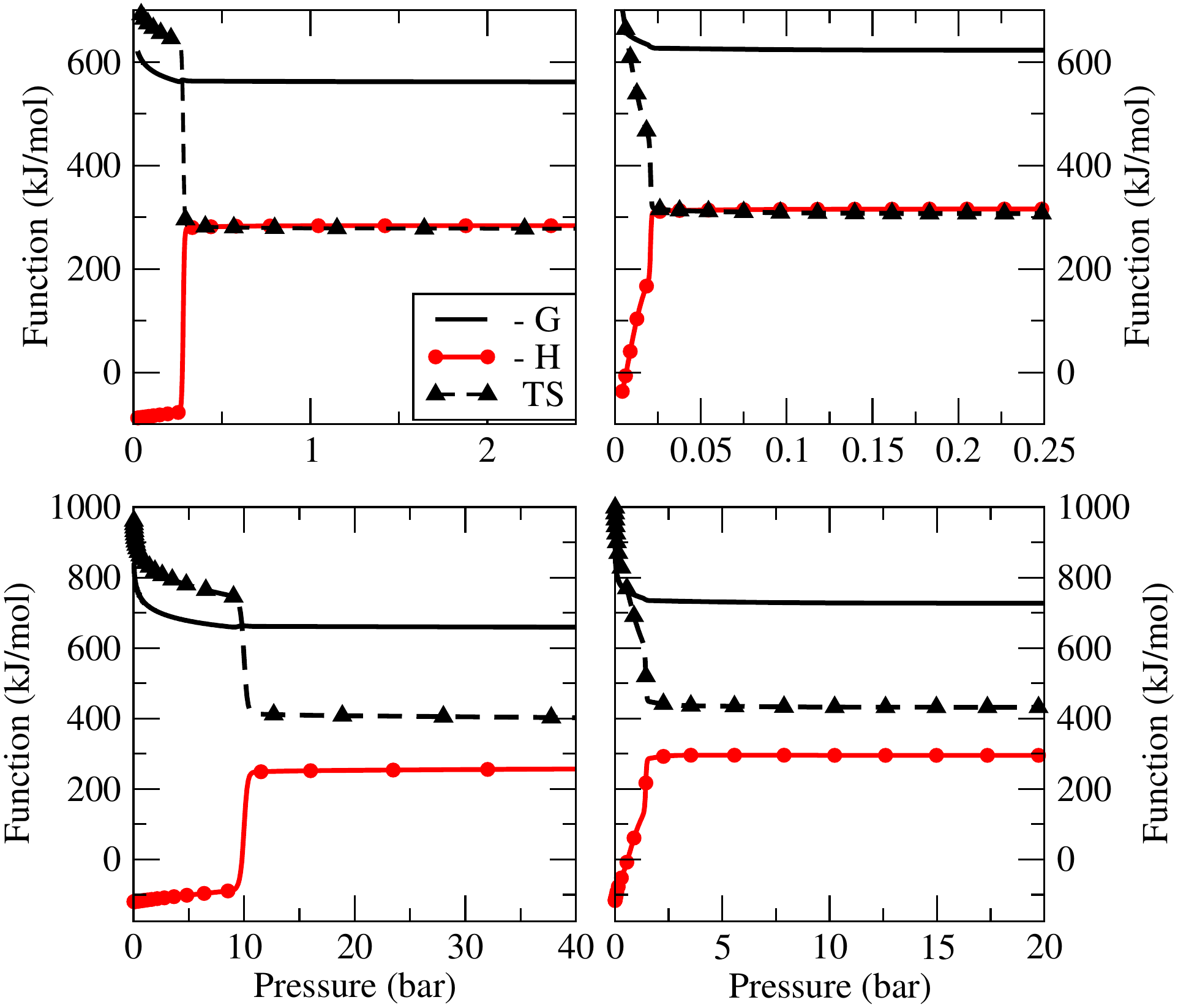}
\end{center}
\caption{DESGRANGES-DELHOMMELLE}
\label{Fig9}
\end{figure}

The partition functions, provided by the EWL approach for the two models, can be used to predict all thermodynamic properties of adsorption, including the Gibbs free energy of adsorption, enthalpy of adsorption and entropy of adsorption~\cite{PartII}. We show on the left panel of Fig.~\ref{Fig9} the predictions for these properties by the CMB-SW model at $T=3000$~K and $T=4000$~K. Fig.~\ref{Fig9} is a plot the opposite of the Gibbs free energy of adsorption, $-G$, against $P$, with $-G$ shown as the sum of the entropic term $TS$ and of the enthalpic term $-H$. For both temperatures, we observe that $-G$ starts to decrease as the pressure increases (for $P$ up to $0.25$~bar at $T=3000$~K). This is mainly due to the slow decrease in entropy, as a result of the slow density increase of the nanoconfined fluid and, to a lesser extent, to the slow increase in $-H$, which also increases with the density of nanoconfined Si. As $P$ further increases (beyond $0.25$~bar at $T=3000$~K), the phase transition from the low-density phase to the high-density phase takes place and results in a steep drop in the entropic term and a steep rise in the enthalpic term. The two terms then reach a plateau, leading to an almost constant value for $-G$ beyond $0.25$~bar. A similar behavior is observed at higher temperature, as shown at the bottom of the left panel for $T=4000$~K. However, while the entropic and enthalpic terms take similar values at $T=3000$~K in the high pressure regime, the entropic term increases with $T$ and becomes greater than the enthalpic term in the high pressure regime. The predictions obtained with the QMB-L model are shown on the right panel of Fig.~\ref{Fig9}. The Gibbs free energy of adsorption for the QMB-L model is shifted by about 10~\% with respect to the CMB-SW model and the transition low-density$\to$~high-density occurs at a lower pressure for the QMB-L model than for the CMB-SW model. However, there is a very good agreement between the predictions from the two models for the relative magnitude of the entropic and enthalpic contributions both at $T=3000$~K and $T=4000$~K. For instance, at $T=3000$~K, $-H$ and $TS$ are found to be the same for both the CMB-SW and the QMB-L force fields in the plateau  region. These sets of results confirm that the CMB-SW and QMB-L force fields, which rely on dramatically different approaches to model many-body effects, both manage to capture the essential features of the thermodynamics of nanoconfined Si.

\section{Conclusions}

We develop the Expanded Wang-Landau method for systems modeled with a tight-binding Hamiltonian and apply the resulting method to determine the partition function and thus all thermodynamic properties of the fluid phases of Si. We consider different strategies, either classical or quantum, to take into account many-body effects in the fluid phases of Silicon. These include classical many-body force fields, which mimic the tetrahedral organization in crystalline Si through a repulsive 3-body term that favoring the ideal tetrahedral angle in the CMB-SW model, as well as quantum many-body tight binding models, which explicitly calculate the overlap between the 4 valence orbitals of neighboring atoms as in the QMB-T and QMB-L models. The EWL results show that the grand-canonical partition function is very sensitive to the strategy (CMB or QMB) chosen to model many-body effects. In particular, the chemical potential for which the system undergoes a vapor$\to$liquid transition, both in the bulk and under nanoconfinement, is predicted to be $10-15$~\% higher for CMB models than for QMB models. When subjected to nanoconfinement, the phase diagram was shown to undergo a dramatic change, with e.g. at $6000$~K a decrease in liquid densities by about $50$~\% for both CMB and QMB models and an increase in vapor densities between $90$~\% (CMB) and $170$\% (QMB). The results obtained in this work also allow us to rank the performance of the various models on the basis of their ability to predict the available experimental data. In particular, the CMB-SW and the QMB-L models are found to yield the results that are the closest to the experimental data for the vapor pressure. With respect to the CMB-SW model, the QMB-L model has the additional advantage of providing an insight into the dependence, upon the thermodynamic conditions, of the electronic properties of the fluid phases of Silicon.\\

{\bf Acknowledgements}

Partial funding for this research was provided by NSF through CAREER award DMR-1052808.

\pagebreak
{\bf List of figures and captions}\\

FIG. 1: Comparison of fractional-full interactions for different values of the coupling parameter $\xi$ to the full-full interaction for the Lenosky model. (a) (Top) Repulsive pair potential and (Bottom) TB matrix elements $h_{ss\sigma}$ (dashed lines) and $h_{sp\sigma}$ (solid lines), and (b) Dimer potential energy.\\

FIG. 2:  Si at $T=5000$~K. (a) Logarithm of the grand-canonical partition function $\Theta (\mu,V,T)$ of Si for the classical force fields, with the results for the SW model shown in black and the results for the Tersoff model shown in red, and for the tight-binding models, with the results for the Kwon model shown in green and the results for the Lenosky model shown in blue. (b) Logarithm of the function $Q(N,V,T)$ (same legend as in (a)).\\

FIG. 3: Density probability $p(\rho)$ obtained with the EWL method for the Lenosky model. (a) 3D-plot showing the results obtained at coexistence as a function of the density and temperature, with the peak corresponding to the vapor phase on the left and the peak corresponding to the liquid phase on the right. (b) Probabilities are shown using a logarithmic scale for the density in the case of the vapor phase (left), and using a linear scale in the case of the liquid phase (right).\\

FIG. 4: Vapor-liquid equilibrium properties for Silicon. (a) Densities of the two fluid phases at coexistence: EWL results are shown for the SW model (circles), for the Tersoff model (diamonds), for the Kwon model (triangles up) and for the Lenosky model (squares), and compared to previous work using GEMC simulations for the SW model~\cite{Honda} (crosses). (b) Vapor pressure results obtained using the EWL method for the 4 models (same symbols as in (a)). The inset shows a comparison in the low-temperature range to the experimental data~\cite{Desai} (solid black line with stars).\\

FIG. 5: Electronic density of states for the Kwon model (left) and the Lenosky model (right) for the liquid phase at coexistence. Results obtained at 3000~K are shown in black, while results obtained at 6000~K are shown in red.\\

FIG. 6: (a) Logarithm of the grand-canonical partition function $\Theta (\mu,V,T)$ for nanoconfined Si. Results for the SW model are shown in black for $3000$~K (solid line) and for $5000$~K (dashed line). Results for the Lenosky model are shown in red for $3000$~K (solid line with filled squares) and for $5000$~K (dashed line with open squares). (b) Logarithm of the function $Q(N,V,T)$ (same legend as in (a)).\\

FIG. 7: Adsorption of Silicon in a slit nanopore for the SW model. (Top) Adsorption isotherms showing the variation of the number of Si atoms adsorbed as a function of the chemical potential. (Bottom) Phase diagram for nanoconfined Si (in red), with a comparison to the bulk (in black).\\

FIG. 8: Adsorption of Silicon in a slit nanopore for the Lenosky model. (Top) Adsorption isotherms showing the variation of the number of Si atoms adsorbed as a function of the chemical potential. (Bottom) Vapor liquid phase equilibria for nanoconfined Si (in red), with a comparison to the bulk (in black).\\

FIG. 9: Adsorption of Silicon in a slit nanopore. The left panel shows the results for the SW model at $3000$~K (top) and $4000$~K (bottom) for the Gibbs free energy of adsorption (black line), the entropy of adsorption (black triangles up with a dashed line) and the enthalpy of adsorption (red circles with a solid line). The right panel shows the results for the Lenosky model at $3000$~K (top) and $4000$~K (bottom), with the same legend as for the left panel.\\

\end{document}